\documentclass[prl,twocolumn,floatfix,
amsmath,amssymb,aps,superscriptaddress]{revtex4-1}
\usepackage{dcolumn,amsmath}
\usepackage{graphicx}
\graphicspath{ {./Images/} }
\usepackage{bm} 
\usepackage{hyperref}     

\begin{document}

\title{Many body study of the Bohr-Weisskopf effect in the thallium atom}

\author{S.D. Prosnyak} \email{prosnyak.sergey@yandex.ru}
\affiliation{National Research Centre Kurchatov Institute B.P. 
Konstantinov Petersburg Nuclear Physics Institute, Gatchina, 
Leningrad District 188300, Russia}
\affiliation{Saint Petersburg State University, 7/9
Universitetskaya nab., St. Petersburg, 199034 Russia}
\author{D.E. Maison}\email{daniel.majson@mail.ru}
\affiliation{National Research Centre Kurchatov Institute B.P. 
Konstantinov Petersburg Nuclear Physics Institute, Gatchina, 
Leningrad District 188300, Russia}
\affiliation{Saint Petersburg State University, 7/9
Universitetskaya nab., St. Petersburg, 199034 Russia}
\author{L.V. Skripnikov}\email{leonidos239@gmail.com}
\affiliation{National Research Centre Kurchatov Institute B.P. 
Konstantinov Petersburg Nuclear Physics Institute, Gatchina, 
Leningrad District 188300, Russia}
\affiliation{Saint Petersburg State University, 7/9
Universitetskaya nab., St. Petersburg, 199034 Russia}

\homepage{http://www.qchem.pnpi.spb.ru}

\date{07.03.2019}

\begin{abstract}
We report the relativistic coupled cluster study of the hyperfine structure and effect of the nuclear magnetization distribution (Bohr-Weisskopf effect) in the $6P_{1/2}$, $6P_{3/2}$ and $7S_{1/2}$ states of several Tl isotopes. It is shown that the Gaussian basis set can be used in such electronic structure calculations and provide a good accuracy. For the ground electronic state of the neutral Tl atom achieved uncertainty for the hyperfine structure constant is smaller than 1\%. A strong Bohr-Weisskopf correction (about 16\%) was found for the $6P_{3/2}$ state which can be of interest for the nuclear structure theory. Basing on the theoretical treatment as well as the available experimental data nuclear magnetic moments of short lived ${}^{191}$Tl and ${}^{193}$Tl isotopes were also predicted.
\end{abstract}

\maketitle

\section{Introduction}
Methods of modern spectroscopy make it possible to measure hyperfine splittings in electronic states of atoms and molecules with high accuracy, resulting in a large amount of data for the analysis and interpretation \cite{barzakh2013changes, barzakh2017changes, beiersdorfer2001hyperfine,schmidt2018nuclear,Petrov:13,Mawhorter:2011}. These data are important not only for the nuclear theory but also for the development and testing methods of precise electronic structure calculations. Such methods are required in the field of searching for the effects of spatial and time-reversal symmetry violating effects of fundamental interactions in atomic and molecular systems~\cite{GFreview,SafronovaRev:2018,Skripnikov:14c,Skripnikov:17c}. 

To reproduce accurately experimental values of hyperfine splitting energy in heavy atoms with uncertainty of the order of 1\% one has to take into account both relativistic and correlation effects at a very good level of theory. Moreover, one should also take into account nuclear structure contributions (as well as the quantum electrodynamics (QED) corrections~\cite{Ginges:2018}). These are contributions from the distribution of the charge (Breit-Rosental effect) \cite{rosenthal1932isotope, crawford1949mf} and magnetization (Bohr-Weisskopf effect) \cite{bohr1950influence, bohr1951bohr} over the nucleus. For many atomic systems these effects have been extensively studied (for example see Refs.
\cite{maartensson1995magnetic, konovalova2017calculation, gomez2008nuclear, ginges2017ground, dzuba1984va}).

In the point nucleus model the ratio of the hyperfine splittings of two different isotopes is proportional to the ratio of nuclear \emph{g}-factors of the isotopes. However, this is not the case for the real system due to the mentioned effects. Corresponding correction is called a magnetic anomaly:
\begin{equation}
{}^{1}\Delta^{2}=\frac{A_1 \mu_2 I_1}{A_2 \mu_1 I_2}-1,
\end{equation}
where $A_1$ and $A_2$ are hyperfine structure (HFS) constants, $\mu_1$ and $\mu_2$ are nuclear magnetic moments and $I_1$ and $I_2$ are nuclear spins of considered isotopes.

Theoretical value of the anomaly strongly depends on the model of the magnetization distribution. However, as was noted previously (e.g. in  Refs.~\cite{0954-3899-37-11-113101,schmidt2018nuclear,barzakh2012hyperfine}) the ratio of anomalies for different electronic states has a very small model dependence. This feature has been employed here to predict nuclear magnetic moments of short lived ${}^{191}$Tl and ${}^{193}$Tl isotopes. 


In the present paper relativistic coupled cluster calculations of the hypefine structure constants and magnetic anomaly in the thallium atom for several electronic states are performed. We show that for this case it is possible to use a simple model of the magnetization distribution in Tl isotopes and fix the parameter of the model from the available experimental data and use it for further predictions. We show that for such electronic structure calculations one can use Gaussian basis sets which implies a direct generalization to similar molecular problems. 

\section{Theory}

For both considered stable isotopes of thallium ($^{203}$Tl and $^{205}$Tl) the total nuclear moment is $1/2$. We consider a simple one-particle model in which the nuclear magnetization is due to single spherically distributed valence nucleon (proton) with the nuclear spin equal to $1/2$ and zero orbital moment. The hyperfine interaction in the case of point magnetic dipole moment is given by the following Hamiltonian:
\begin{equation} 
\label{hfs1}
h^{\rm HFS}=\frac{1}{c}\frac{\mathbf{\mu}\cdot[\mathbf{r}_{el}\times \bm{\alpha}]}{r_{el}^3},
\end{equation}
where $\bm{\alpha}$ are Dirac matrices and $\mathbf{r}_{el}$ is the electron radius-vector. In the model of the uniformly magnetized ball with radius $R_m$ this interaction inside the nucleus modifies to the following form~\cite{bohr1950influence}:
\begin{equation}
\label{hfs2}
\frac{1}{c}\frac{\mathbf{\mu}\cdot[\mathbf{r}_{el}\times \bm{\alpha}]}{R_m^3}.
\end{equation}
Outside the nucleus the expression for the interaction coincides with that of the point dipole given by Eq. (\ref{hfs1}). In this paper we use the root-mean-square radius $r$ associated with the radius of the ball $R$ by $R=(\frac{5}{3} r)^{1/2}$.

For the hyperfine structure constant one often uses the following parametrization:
\begin{equation}
    A=A_0(1-\delta)(1-\epsilon),
\end{equation}
where $A_0$ is the hyperfine structure constant corresponding to the point nucleus, $\delta$ is the Breit-Rosenthal correction and $\epsilon$ is the Bohr-Weisskopf (BW) correction. The former takes into account the finite charge distribution and is about 10\% for heavy atoms like Tl \cite{shabaev1994hyperfine}. The latter concerns finite magnetization distribution and usually smaller than $\delta$.

In the relativistic correlation calculations of the neutral atoms it is more practical to use the finite charge distribution model from the beginning. In the present paper the Gaussian charge distribution has been employed. Thus, we do not introduce the Breit-Rosenthal correction. In this case the Bohr-Weisskopf correction $\epsilon$ is a function of both nuclear charge and magnetic radii: $\epsilon = \epsilon(r_c, r_m)$.

For hydrogen-like ions one can use the following analytic expression for the hyperfine splitting energy~\cite{shabaev1997ground}:

\begin{equation} \label{eq1}
\begin{split}
\Delta E & = \frac{4}{3}\alpha (\alpha Z)^3 
    \frac{\mu}{\mu _N} \frac{m}{m _p}
    \frac{2I+1}{2I}mc^2 \\
 & \times \big (A(\alpha Z)(1-\delta)
(1-\varepsilon)+x_{rad} \big)
\end{split}
\end{equation}
where $\alpha$ is the fine-structure constant, $Z$ is the nuclear charge, $m_p$ is the proton mass, $m$ is the electron mass and $x_{rad}$ is the correction due to QED effects. Besides, the following analytic behavior for the Breit-Rosenthal correction was obtained~\cite{ionesco1960nuclear}:
\begin{equation}
\label{dep}
\delta = b_N \cdot R_N^{2\gamma -1} 
\text{   ,   }
\gamma = \sqrt{\kappa^2-(\alpha Z)^2}.
\end{equation}
Here $b_N$ is a constant independent of the nuclear structure, $\kappa$ is the relativistic quantum number. 
This expression can be used to test numerical approaches.

Finally, in Ref.~\cite{atoms6030039} the following parametrization of the hyperfine structure constant and Bohr-Weisskopf correction has been used:
\begin{eqnarray}
    \label{d_nuc_dep}
    A=A_0(1-(b_N+b_M d_{nuc}R_N^{2\gamma - 1})),\\
    \epsilon(R_N, d_{nuc})=b_M d_{nuc}R_N^{2\gamma - 1},
\end{eqnarray}
where $b_M$ is the electronic factor independent of the nuclear radius and structure and $d_{nuc}$ is a factor which depends only on the nuclear spin and configuration.

\section{Calculation details}


In the calculations of the hyperfine splittings in the hydrogen-like thallium, the values of the QED corrections from Ref.~\cite{shabaev1997ground} were used. The values of the nuclear charge radii were taken from Ref.~\cite{ANGELI201369}. The nuclear magnetic moments of $^{203}$Tl and $^{205}$Tl isotopes were taken from Ref.~\cite{stone2005table}.


For test calculations of the hydrogen-like Tl three basis sets were used: CVDZ~\cite{Dyall:06,Dyall:98} which consists of $24s, 20p, 14d, 9f$ uncontracted Gaussian functions with a maximum $s$-type exponent parameter equal to $5.8 \cdot 10^7$, GEOM1 which consists of $50s$ functions with exponent parameters forming the geometric progression, where a common ratio is equal to $1.8$ and the largest term is $5 \cdot 10^8$, as well as GEOM2, which differs from the previous one only in that the first term is $5 \cdot 10^9$.

In the calculations of the neutral thallium atoms the QED effects were not taken into account. The main all-electron (81e) correlation calculations were performed within the coupled cluster with single, double and perturbative triple cluster amplitudes, CCSD(T), method \cite{bartlett2007coupled} using the Dirac-Coulomb Hamiltonian.
In the calculation the the AAE4Z basis set~\cite{Dyall:12} augmented with one $h$ and one $i$ functions was used. It includes $35s, 32p, 22d, 16f, 10g, 5h, 2i$. This basis set is called LBas below. For the calculation virtual orbitals were truncated at the energy of 10000 Hartree. Importance of the high energy cutoff for properties dependent on the behaviour of the wave function close to a heavy atom nucleus were analyzed in Ref.~\cite{Skripnikov:17a}. Besides, the basis set correction has been calculated. For this we have considered the extended basis set LBasExt which includes $44s, 40p, 31d, 24f, 15g, 9h, 8i$ basis functions. This correction has been calculated within the CCSD(T) method with frozen $1s..3d$ electrons. Virtual orbitals were truncated at the energy of 150 Hartree. 
To test convergence with respect to electron correlation effects we have used the SBas basis set which consists of $30s, 26p, 15d, 9f$ functions and equals to the CVDZ~\cite{Dyall:06,Dyall:98} basis set augmented by diffuse functions.
To calculate the correlation correction $1s-3d$ electrons were excluded from the correlation treatment. In such a way correlation effects up to the level of the coupled cluster with single, double, triple and perturbative quadruple amplitudes, CCSDT(Q) method were considered. Contribution of the Gaunt interaction was calculated within the SBas basis set using the CCSD(T) method. In this calculation all electrons were correlated.

For a number of intermediate all-electron correlation calculations we also used the MBas basis set consisting of $31s, 27p, 18d, 13f, 4g, 1h$ functions and corresponding to the AAETZ basis set~\cite{Dyall:06}.

For the relativistic coupled cluster calculations the {\sc dirac15} \cite{DIRAC15} and {\sc mrcc} codes \cite{MRCC2013,Kallay:1,Kallay:2} were used. The code developed in Ref. \cite{Skripnikov:16b} has been used to compute point dipole magnetic HFS constant integrals. To treat the Gaunt interaction contributions we used the code developed in Ref.~\cite{Maison:2019}. The code to compute the Bohr-Weisskopf correction was developed in the present paper.

\section{Results and discussion}
\subsection{Hydrogen-like thallium \texorpdfstring{$1S_{1/2}$}{1S 1/2} }

\begin{figure}[ht]
\begin{center}
\includegraphics[width=1.0\linewidth]{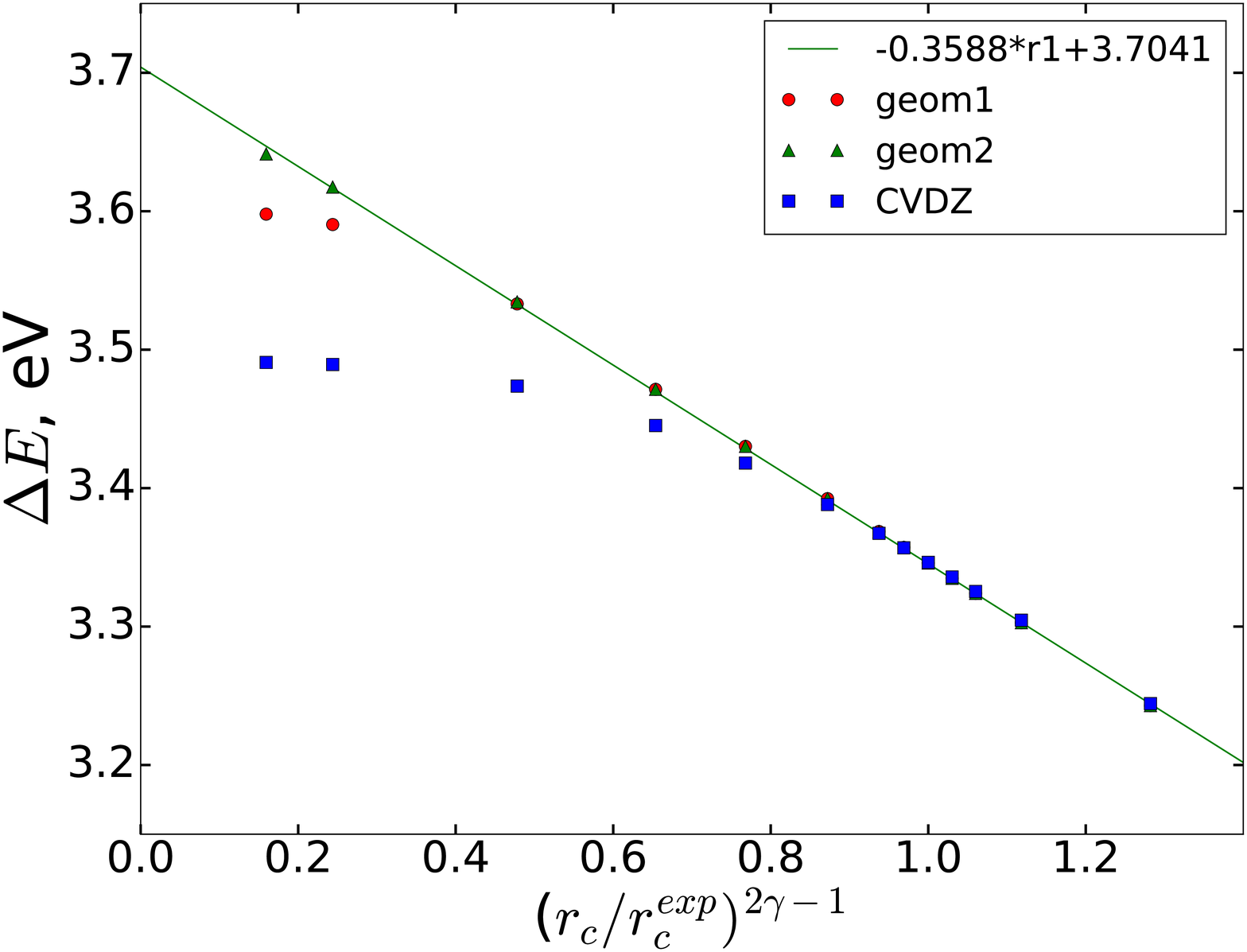}
\end{center}
\caption{Calculated dependence of the hyperfine splitting, $\Delta E$, of the ground electronic state of the hydrogen like ${}^{205}$Tl on $r_c^{2\gamma-1}$ in the point  nuclear magnetic dipole moment model in different basis sets (see text). $r_c$ is the nuclear charge radius.}
\label{figure:1}
\end{figure}

Fig. \ref{figure:1} shows calculated dependence of the hyperfine splitting of the ground electronic state of the hydrogen like ${}^{205}$Tl in the point nuclear magnetic dipole moment model  (without QED correction) on $r_c^{2\gamma-1}$ in different basis sets. One can see that this calculated dependence is in a very good agreement with the analytical dependence given by Eq.~(\ref{dep}). Extrapolated value of the hyperfine splitting for the point like nucleus (3.7041 eV) almost coincides with the analytical value obtained within Eq.~(\ref{eq1}) (3.7042 eV) for the GEOM2 basis set.

\begin{figure}[ht]
\begin{center}
\includegraphics[width=1.0\linewidth]{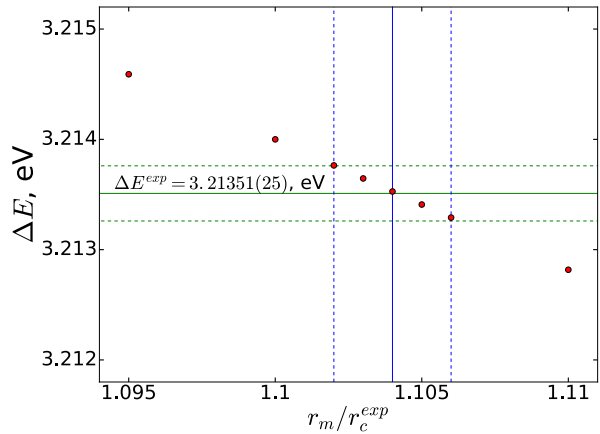}
\end{center}
\caption{
Calculated dependence of the hyperfine splitting ($\Delta E$) of the ground electronic state of the hydrogen like ${}^{203}$Tl on the ratio $r_m / r_c^{exp}$ of the magnetic radius $r_m$ and experimental charge radius $r_c^{exp}$. Solid and dashed horizontal lines show the experimental energy splitting value with the corresponding uncertainty from Ref.~\cite{beiersdorfer2001hyperfine}. Vertical lines show the extracted magnetic radius with the corresponding uncertainty.}
\label{figure:2}
\end{figure}

\begin{figure}[ht]
\begin{center}
\includegraphics[width=1.0\linewidth]{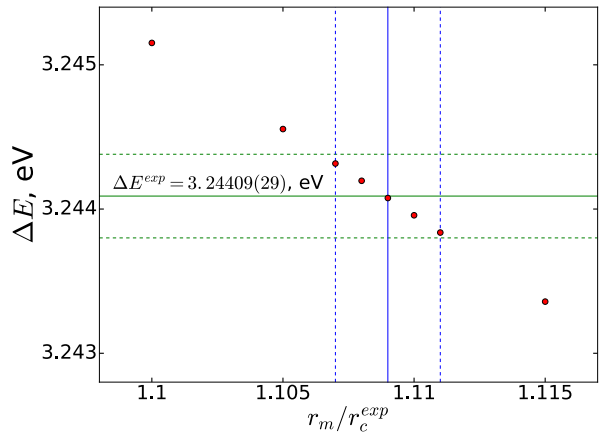}
\end{center}
\caption{
Calculated dependence of the hyperfine splitting ($\Delta E$) of the ground electronic state of the hydrogen like ${}^{205}$Tl on the ratio $r_m / r_c^{exp}$ of the magnetic radius $r_m$ and experimental charge radius $r_c^{exp}$. Solid and dashed horizontal lines show the experimental energy splitting value with the corresponding uncertainty from Ref.~\cite{beiersdorfer2001hyperfine}. Vertical lines show the extracted magnetic radius with the corresponding uncertainty.}

\label{figure:3}
\end{figure}

Figs. \ref{figure:2} and \ref{figure:3} give calculated dependence of hyperfine splittings of the ground electronic state of the hydrogen like $^{203}$Tl and $^{205}$Tl on magnetic radii. Horizontal lines show the experimental energy splitting with the corresponding uncertainty taken from Ref.~\cite{beiersdorfer2001hyperfine}. From these data it is possible to fix magnetic radii for the used model of the magnetization distribution. For $^{205}$Tl one obtains $r_m / r_c^{exp} = $ 1.109(2) and for $^{203}$Tl $r_m / r_c^{exp} = $ 1.104(2).

Combining theoretical and experimental data, the coefficients $d_{nuc}=1.17$ for ${}^{203}$Tl and $d_{nuc}=1.18$ for ${}^{205}$Tl were obtained for the parametrization given by Eq.~(\ref{d_nuc_dep}).

\subsection{Neutral thallium 
\texorpdfstring{${}^{205}$Tl}{Tl 205} in 
\texorpdfstring{$6P_{1/2}$}{6P 1/2} state}


Table \ref{table:2} gives calculated values of the ${}^{205}$Tl hyperfine structure constant for the $6P_{1/2}$ state for a number of magnetic radii. The last column gives values for $r_m/r_c^{exp}=1.11$ which is close to the value obtained from the analysis given above for the hyperfine splitting in the hydrogen like ${}^{205}$Tl.

\begin{table}[ht]
\centering
\begin{tabular}{lrrrr}
\hline
\hline
$r_m/r_c^{exp}$ & 0 & 1 & 1.1 & 1.11 \\
\hline
DHF &~ 18805 &~ 18681 &~ 18660 &~ 18658\\
CCSD & 21965 & 21807 & 21781 & 21778\\
CCSD(T) & 21524 & 21372 & 21347 & 21345\\
\hline
~+Basis corr. & -21 & -21 & -21 & -21 \\
~+CCSDT-CCSD(T) & +73 & +73 & +73 & +73 \\
~+CCSDT(Q)-CCSDT & -5 & -5 & -5 & -5\\
~+Gaunt & -83 & -83 & -83 & -83 \\
\hline
Total & 21488 & 21337 & 21312 & 21309\\
\hline
\hline
\end{tabular}
\caption{Calculated values of the hyperfine structure constant of the $6P_{1/2}$ state of ${}^{205}$Tl (in MHz) at different levels of theory.}
\label{table:2}
\end{table}



The final value for the HFS constant with accounting for the Bohr-Weisskopf correction is 21309 MHz and is in a very good agreement with the experimental value 21310.8 MHz \cite{lurio1956hfs} and previous studies (see Table~\ref{tableCompare}). One can estimate the theoretical uncertainty of the calculated HFS constant to be smaller than 1\%.

\subsection{Neutral thallium 
\texorpdfstring{${}^{205}$Tl}{Tl 205} in 
\texorpdfstring{$6P_{3/2}$}{6P 3/2} state}


Table \ref{table:5} gives calculated values of the HFS constant for the $6P_{3/2}$ state of the ${}^{205}$Tl atom. One can see that correlation effects dramatically contribute to the constant. This has also been noted in previous studies of this state \cite{konovalova2017calculation, maartensson1995magnetic}. Interestingly that even quadruple cluster amplitudes give non-negligible relative contribution to the HFS constant.

\begin{table}[ht]
\centering
\begin{tabular}{lrrrr}
\hline
\hline
$r_m/r_c^{exp}$ & 0 & 1 & 1.1 & 1.11 \\
\hline
DHF &~ 1415 &~ 1415 &~ 1415 &~ 1415\\
CCSD & 6 & 40 & 46 & 47\\
CCSD(T) & 244 & 273 & 278 & 278\\
\hline
~+Basis corr. & +4 & +4 & +4 & +4\\
~+CCSDT-CCSD(T) & -49 & -49 & -49 & -49 \\
~+CCSDT(Q)-CCSDT & +13 & +13 & +13 & +13 \\
~+Gaunt & +1 & +1 & +1 & +1 \\
\hline
Total & 214 & 243& 248 & 248\\
\hline
\hline
\end{tabular}
\caption{Calculated values of the hyperfine structure constant of the $6P_{3/2}$ state of ${}^{205}$Tl (in MHz) at different levels of theory}
\label{table:5}
\end{table}


Calculated value of the BW correction to the HFS constant of the $6P_{3/2}$ state of ${}^{205}$Tl is $-$16\%. It has an opposite sign with respect to the BW correction to the HFS constant of the $6P_{1/2}$ (see Table \ref{table:6}).

\begin{table}[ht]
\centering
\begin{tabular}{lrrr}
\hline
\hline
$r_m/r_c^{exp}$ & 1 & 1.1 & 1.11 \\
\hline
$6P_{1/2}$  &~ 0.0070 & 0.0082 &~ 0.0083\\
$6P_{3/2}$ & -0.14 & -0.16 & -0.16\\
\hline
\hline
\end{tabular}
\caption{
Calculated values of the Bohr-Weisskopf correction to the hyperfine structure constants of the $6P_{1/2}$ and $6P_{3/2}$ states of ${}^{205}$Tl for different values of the $r_m/r_c^{exp}$ ratio.
}
\label{table:6}
\end{table}


\begin{table}[]
    \centering
\begin{tabular}{lrr}
    \hline
    \hline
    Author, Ref. &~ $6P_{1/2}$ &~ $6P_{3/2}$ \\
    \hline
    Kozlov et al. \cite{kozlov2001parity} & 21663 & 248 \\
    Safronova et al. \cite{safronova2005excitation} & 21390 & 353 \\
    M{\aa}rtensson-Pendrill \cite{maartensson1995magnetic} & 20860 & 256 \\
    This work & 21488 & 214 \\
    \hline
    \hline
\end{tabular}
    \caption{Total values of hyperfine structure constants in the point nuclear magnetic dipole moment model for ${}^{205}$Tl (in MHz) in comparison with previous studies.}
    \label{tableCompare}
\end{table}

Obtained value of the HFS constant is in a reasonable agreement with the experimental value 265 MHz \cite{gould1956hfs}. Theoretical uncertainty of the final value can be estimated as 10\%. Note, however, that it corresponds to about 2\% of the total correlation contribution -- compare the final value with the Dirac-Hartree Fock (DHF) value.

\section{Hyperfine anomaly}

\begin{figure}[ht]
\begin{center}
\includegraphics[width=1.0\linewidth]{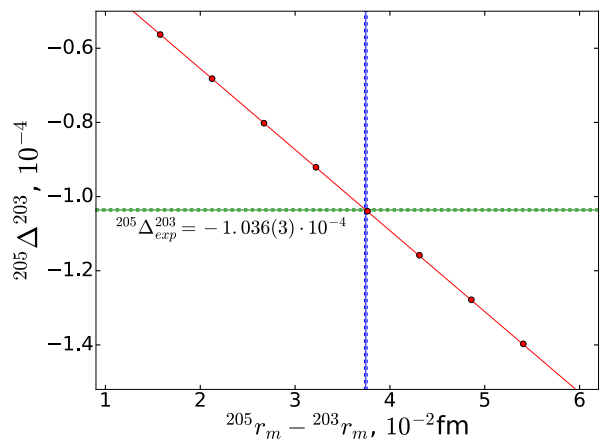}
\end{center}
\caption{
Calculated dependence of the $6P_{1/2}$ hyperfine anomaly ${}^{205}\Delta^{203}$ on the difference of magnetic radii ${}^{205}r_m-{}^{203}r_m$. Solid and dashed horizontal lines indicate the experimental value and its uncertainty; calculated values are given by circles and vertical lines give the value of the magnetic radii difference and its uncertainty.}
\label{figure:4}
\end{figure}

\begin{figure}[ht]
\begin{center}
\includegraphics[width=1.0\linewidth]{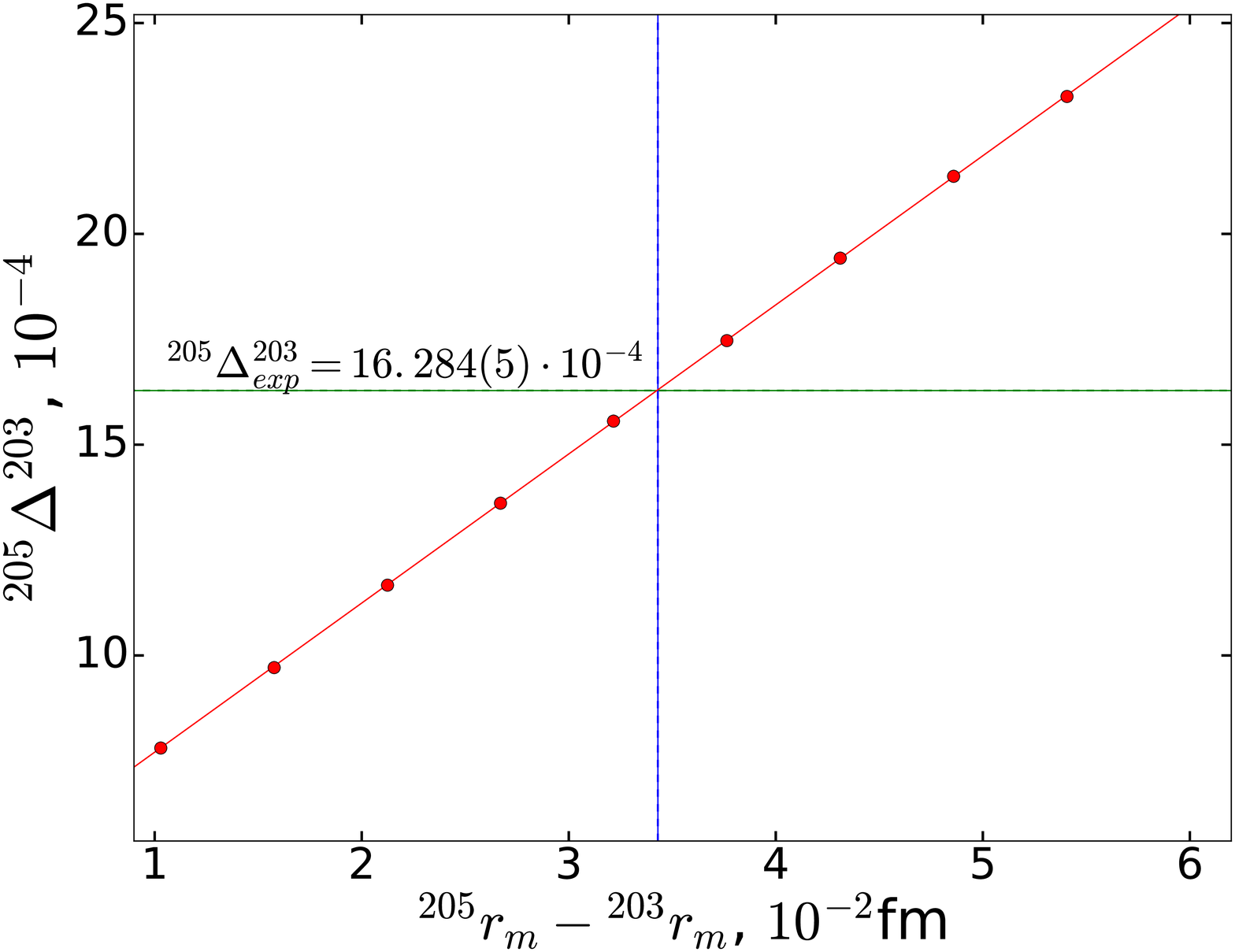}
\end{center}
\caption{
Calculated dependence of the $6P_{3/2}$ hyperfine anomaly ${}^{205}\Delta^{203}$ on the difference of magnetic radii ${}^{205}r_m-{}^{203}r_m$. Solid and dashed horizontal lines indicate the experimental value and its uncertainty; calculated values are given by circles and vertical lines give the value of the magnetic radii difference and its uncertainty.}
\label{figure:5}
\end{figure}

\begin{figure}[ht]
\begin{center}
\includegraphics[width=1.0\linewidth]{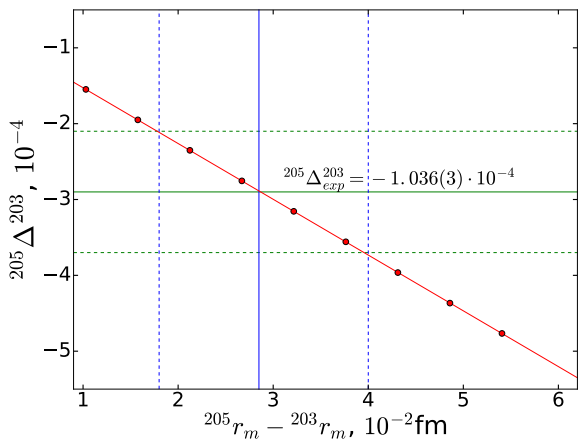}
\end{center}
\caption{
Calculated dependence of the $7S_{1/2}$ hyperfine anomaly ${}^{205}\Delta^{203}$ on the difference of magnetic radii ${}^{205}r_m-{}^{203}r_m$. Solid and dashed horizontal lines indicate the experimental value and its uncertainty; calculated values are given by circles and vertical lines give the value of the magnetic radii difference and its uncertainty.}
\label{figure:6}
\end{figure}

Magnetic moments of $^{203}$Tl and $^{205}$Tl are known with a good accuracy \cite{stone2005table}. Values of HFS constants of the $6P_{1/2}$, $6P_{3/2}$ and $7S_{1/2}$ states have been measured precisely in Refs.~\cite{lurio1956hfs, gould1956hfs, chen2012absolute}. Thus, experimental values of the magnetic anomalies ${}^{205}\Delta^{203}$ for these states are also known with a high precision. Figs. \ref{figure:4}, \ref{figure:5} and \ref{figure:6} show
calculated dependence of the values of anomalies for these states on the difference of magnetic radii ${}^{205}r_m-{}^{203}r_m$. In these calculations charge radii of $^{203}$Tl and $^{205}$Tl were set to experimental values. 
Calculations were performed within the CCSD(T) method in the MBas basis set.
In Figs. \ref{figure:4}, \ref{figure:5} and \ref{figure:6} solid and dashed horizontal lines show the experimental value and its uncertainty. By considering the intersection of the calculated (without treatment of the QED effects) dependence (which is approximated by linear functions) with horizontal dashed lines one obtains the difference of magnetic radii ${}^{205}r_m-{}^{203}r_m$ and its uncertainty in the model under consideration.

One can see from Figs. \ref{figure:4} and \ref{figure:5} that the differences extracted from the data for the $6P_{1/2}$ and $6P_{3/2}$ states agree within 10\% which confirms the applicability of the model under consideration. They are also in a good agreement with the difference obtained from the data for the $7S_{1/2}$ state of the neutral Tl as well as from the data for the hydrogen like Tl above --- within the experimental uncertainty for these systems.




\section{Magnetic moments of short lived isotopes}

Magnetic anomalies can be used to determine the value of the nuclear magnetic moment of the short lived isotope (for example see Refs.~\cite{0954-3899-37-11-113101,schmidt2018nuclear,barzakh2012hyperfine}).
For this one should know the nuclear magnetic moment value of the stable isotope as well as HFS constants ($A[a]$ and $A[b]$) for two electronic states ($a$ and $b$) of this isotope and the short lived isotope. Consider isotopes 1 (${}^{205}$Tl, $I=1/2$) and 2 (${}^{193}$Tl, $I=9/2$). The latter is unstable.
From the experimental data \cite{barzakh2012hyperfine, lurio1956hfs, chen2012absolute}  one obtains:
\begin{equation}
    {}^1\theta^2[a,b]=
    \frac{A_1[a]}{A_2[b]}\frac{A_2[a]}{A_1[b]}-1=
    -0.013(7).
\end{equation}
Here $a=7S_{1/2}$, $b=6P_{1/2}$. Calculated value of the ratio of magnetic anomalies for these states is 
${}^1k^2[a,b]={}^{1}\Delta^{2}[a]/{}^{1}\Delta^{2}[b]=3.4(2)$. Such ratio depends only slightly on the nuclear magnetization distribution model \cite{0954-3899-37-11-113101,schmidt2018nuclear}. Now one obtains for the nuclear magnetic moment $\mu_2$ of the isotope 2:
\begin{equation}
    \mu_2=\mu_1\cdot\frac{A_2[b]}{A_1[b]}\cdot\frac{I_2}{I_1}
\cdot(1+{}^{1}\Delta^{2}[b])=3.84(3).
\end{equation}

Using the same method and experimental data from Refs.~\cite{barzakh2012hyperfine, lurio1956hfs, chen2012absolute} one obtains the nuclear magnetic moment value of the ${}^{191}$Tl isotope with nuclear spin $I=9/2$: $\mu_{191}=+3.79(2)$.
Obtained values of $\mu_{191}$ and $\mu_{193}$ are in a very good agreement with    Ref.~\cite{barzakh2012hyperfine}: $\mu_{193}=+3.82(3)$ and $\mu_{191}=+3.78(2)$.


\section{Conclusion}

In the present paper the Bohr-Weisskopf effect has been calculated for the $6P_{1/2}$, $6P_{3/2}$ and $7S_{1/2}$ states for several isotopes of the Tl atom. The uniformly magnetized ball model has been tested and used.

It was found that the correlation effects strongly contribute to the HFS constant (they are about 470\% of the final value) as well as to the BW effect for the $6P_{3/2}$ state. BW correction for the $6P_{3/2}$ state was found to be about $-$16\%. Such a significant contribution makes it possible to test models of the nuclear magnetization distribution.

Combining obtained theoretical values of magnetic anomalies and available experimental data nuclear magnetic moments of short-lived ${}^{191}$Tl and ${}^{193}$Tl isotopes were predicted and found to be in a good agreement with the previous study~\cite{barzakh2012hyperfine}.

It was demonstrated that for such calculations the Gaussian basis sets can be used. Thus, the applied method can be extended to the calculation of the BW effect in molecules.

For the calculated value of the HFS constant of the $6P_{1/2}$ state a very good agreement with the experiment and the small theoretical uncertainty has been obtained. A further improvement can be achieved by the treatment of the QED effects contribution.






\section{Acknowledgment}

We are grateful to Prof. M. Kozlov, Prof. I. Mitropolsky and Dr. Yu. Demidov for helpful discussions. Electronic structure calculations were performed at the PIK data center of NRC ``Kurchatov Institute'' -- PNPI.


\end{document}